\documentclass[12pt]{article}
\usepackage{graphicx}
\usepackage{lscape}
\usepackage{citesort}
\usepackage{amssymb}
\usepackage{appendix}
\usepackage{multirow}

\begin{document}
\def\qq{\langle \bar q q \rangle}
\def\uu{\langle \bar u u \rangle}
\def\dd{\langle \bar d d \rangle}
\def\sp{\langle \bar s s \rangle}
\def\GG{\langle g_s^2 G^2 \rangle}
\def\Tr{\mbox{Tr}}
\def\figt#1#2#3{
        \begin{figure}
        $\left. \right.$
        \vspace*{-2cm}
        \begin{center}
        \includegraphics[width=10cm]{#1}
        \end{center}
        \vspace*{-0.2cm}
        \caption{#3}
        \label{#2}
        \end{figure}
    }

\def\figb#1#2#3{
        \begin{figure}
        $\left. \right.$
        \vspace*{-1cm}
        \begin{center}
        \includegraphics[width=10cm]{#1}
        \end{center}
        \vspace*{-0.2cm}
        \caption{#3}
        \label{#2}
        \end{figure}
                }

\def\ds{\displaystyle}
\def\beq{\begin{equation}}
\def\eeq{\end{equation}}
\def\bea{\begin{eqnarray}}
\def\eea{\end{eqnarray}}
\def\beeq{\begin{eqnarray}}
\def\eeeq{\end{eqnarray}}
\def\ve{\vert}
\def\vel{\left|}
\def\ver{\right|}
\def\nnb{\nonumber}
\def\ga{\left(}
\def\dr{\right)}
\def\aga{\left\{}
\def\adr{\right\}}
\def\lla{\left<}
\def\rra{\right>}
\def\rar{\rightarrow}
\def\lrar{\leftrightarrow}
\def\nnb{\nonumber}
\def\la{\langle}
\def\ra{\rangle}
\def\ba{\begin{array}}
\def\ea{\end{array}}
\def\tr{\mbox{Tr}}
\def\ssp{{\Sigma^{*+}}}
\def\sso{{\Sigma^{*0}}}
\def\ssm{{\Sigma^{*-}}}
\def\xis0{{\Xi^{*0}}}
\def\xism{{\Xi^{*-}}}
\def\qs{\la \bar s s \ra}
\def\qu{\la \bar u u \ra}
\def\qd{\la \bar d d \ra}
\def\qq{\la \bar q q \ra}
\def\gGgG{\la g^2 G^2 \ra}
\def\q{\gamma_5 \not\!q}
\def\x{\gamma_5 \not\!x}
\def\g5{\gamma_5}
\def\sb{S_Q^{cf}}
\def\sd{S_d^{be}}
\def\su{S_u^{ad}}
\def\sbp{{S}_Q^{'cf}}
\def\sdp{{S}_d^{'be}}
\def\sup{{S}_u^{'ad}}
\def\ssp{{S}_s^{'??}}

\def\sig{\sigma_{\mu \nu} \gamma_5 p^\mu q^\nu}
\def\fo{f_0(\frac{s_0}{M^2})}
\def\ffi{f_1(\frac{s_0}{M^2})}
\def\fii{f_2(\frac{s_0}{M^2})}
\def\O{{\cal O}}
\def\sl{{\Sigma^0 \Lambda}}
\def\es{\!\!\! &=& \!\!\!}
\def\ap{\!\!\! &\approx& \!\!\!}
\def\md{\!\!\!\! &\mid& \!\!\!\!}
\def\ar{&+& \!\!\!}
\def\ek{&-& \!\!\!}
\def\kek{\!\!\!&-& \!\!\!}
\def\cp{&\times& \!\!\!}
\def\se{\!\!\! &\simeq& \!\!\!}
\def\eqv{&\equiv& \!\!\!}
\def\kpm{&\pm& \!\!\!}
\def\kmp{&\mp& \!\!\!}
\def\mcdot{\!\cdot\!}
\def\erar{&\rightarrow&}
\def\olra{\stackrel{\leftrightarrow}}
\def\ola{\stackrel{\leftarrow}}
\def\ora{\stackrel{\rightarrow}}

\def\simlt{\stackrel{<}{{}_\sim}}
\def\simgt{\stackrel{>}{{}_\sim}}


\title{
         {\Large
                 {\bf
                     Heavy $\chi_{Q_2}$ tensor
                   mesons in QCD
                 }
         }
      }

\author{\vspace{1cm}\\
{\small T. M. Aliev$^a$ \thanks {e-mail:
taliev@metu.edu.tr}~\footnote{permanent address:Institute of
Physics,Baku,Azerbaijan}\,\,, K. Azizi$^b$ \thanks {e-mail:
kazizi@dogus.edu.tr}\,\,, M. Savc{\i}$^a$ \thanks
{e-mail: savci@metu.edu.tr}} \\
{\small $^a$ Physics Department, Middle East Technical University,
06531 Ankara, Turkey} \\
{\small $^b$ Physics Division, Faculty of Arts and
Sciences, Do\u gu\c s University} \\
{\small Ac{\i}badem-Kad{\i}k\"oy, 34722 Istanbul} }
\date{}

\begin{titlepage}
\maketitle
\thispagestyle{empty}

\begin{abstract}
The masses and decay constants of the ground state heavy
$\chi_{Q2}(Q=b,c)$  tensor mesons  are calculated in the framework of the
QCD sum rules approach. The obtained results  on the masses are in good consistency
with the  experimental values. Our predictions on the decay constants can be verified in the future experiments.
\end{abstract}

~~~PACS number(s): 11.55.Hx,  14.40.Pq
\end{titlepage}

\section{Introduction}
During last few years very exiting experimental results are obtained
in the charm and beauty meson and baryon spectroscopies \cite{Amsler}.
Recent CLEO measurements on the two-photon decay rates of the even-parity,
 scalar  $0^{++}, \chi_{b(c)0}$ and tensor $2^{++}, \chi_{b(c)2}$
states (\cite{Amsler,Ecklund} and references therein) were
motivation to investigate the properties of these mesons and their
radiative decays.

In the present work, we calculate the  mass and decay constants of the
ground state heavy bottomonium, $\chi_{b2}(1P)$ and   charmonium,
$\chi_{c2}(1P)$ tensor mesons with $I^G(J^{PC})=0^+(2^{++})$ in the framework
of  the  QCD sum rules approach.
QCD sum rules approach as a non-perturbative approach is one of the most
powerful and applicable tools to hadron physics and can play an important role
in calculation of the characteristic parameters of the hadrons (for details
about this method and some applications see \cite{svz,colangelo}).
Note that the mass and decay constant of the  strange tensor
$K_2^*(1430)$ with quantum numbers $I(J^P)=1/2(2^+)$ have been calculated in
\cite{kazem1} in the same framework. These parameters for
light unflavored tensor mesons have also been calculated in \cite{aliev}.
The obtained results for the masses and decay constants are used in calculation
of the magnetic dipole moments of the light tensor mesons
using the QCD sum rules method in \cite{kazem2}.

The  paper is organized as follows: in  next section,  sum rules for the
mass and decay constant of the ground state heavy quarkonia, $\chi_{Q2}$
tensor mesons are derived in the context of the QCD sum rules method.
Section 3 is devoted to the  numerical analysis of the mass and  decay constants
as well as the comparison of the obtained results on the mass with the
 experimental values.

\section{Theoretical Framework}
In this section, we obtain the sum rules for the mass and decay constant of the heavy
$\chi_{Q2}(1P)$ tensor meson in the framework of the QCD sum rules approach.
For this aim we consider the following correlation function

\begin{eqnarray}\label{correl.func.101}
\Pi _{\mu\nu,\alpha\beta}=i\int
d^{4}xe^{iq(x-y)}{\langle}0\mid {\cal T}[j _{\mu\nu}(x)
\bar j_{\alpha\beta}(y)]\mid  0{\rangle},
\end{eqnarray}
where, $j_{\mu\nu}$ is the interpolating current of the $\chi_{Q2}(1P)$
tensor meson and ${\cal T}$ is the time ordering operator.
The explicit form of the  current $j_{\mu\nu}$ creating the  ground state
heavy tensor  $\chi_{Q2}(1P)$ state with quantum numbers
$I^G(J^{PC})=0^+(2^{++})$ from the vacuum can be written in the following
form:
\begin{eqnarray}\label{tensorcurrent}
j _{\mu\nu}(x)=\frac{i}{2}\left[\bar Q(x) \gamma_{\mu} \olra{\cal D}_{\nu}(x)
Q(x)+\bar Q(x) \gamma_{\nu}  \olra{\cal D}_{\mu}(x) Q(x)\right],
\end{eqnarray}
where $Q$ stands for heavy b or c quark and the $ \olra{\cal D}_{\mu}(x)$
represents the derivative with respect to four-x acting on left and right,
simultaneously. This two-side covariant derivative is defined as:
\begin{eqnarray}\label{derivative}
\olra{\cal D}_{\mu}(x)=\frac{1}{2}\left[\ora{\cal D}_{\mu}(x)-
\ola{\cal D}_{\mu}(x)\right],
\end{eqnarray}
where,
\begin{eqnarray}\label{derivative2}
\overrightarrow{{\cal D}}_{\mu}(x)=\overrightarrow{\partial}_{\mu}(x)-i
\frac{g}{2}\lambda^aA^a_\mu(x),\nonumber\\
\overleftarrow{{\cal D}}_{\mu}(x)=\overleftarrow{\partial}_{\mu}(x)+
i\frac{g}{2}\lambda^aA^a_\mu(x).
\end{eqnarray}
In the above relations, the $\lambda^a$ are the Gell-Mann matrices and
$A^a_\mu(x)$ are the external (vacuum) gluon fields , which can be
expressed directly in terms of the gluon field strength tensor
using the Fock-Schwinger gauge, $x^\mu A^a_\mu(x)=0$, as the following way:
\begin{eqnarray}\label{gluonfield}
A^{a}_{\mu}(x)=\int_{0}^{1}d\alpha \alpha x_{\beta} G_{\beta\mu}^{a}(\alpha x)=
\frac{1}{2}x_{\beta} G_{\beta\mu}^{a}(0)+\frac{1}{3}x_\eta x_\beta {\cal D}_\eta
G_{\beta\mu}^{a}(0)+...
\end{eqnarray}
Since the
current contains derivatives with respect to the space-time so we will
consider the two currents at points x and y. After doing  calculations and applying the derivatives, we will set $y=0$.

It is well known that in the QCD sum rules approach, the correlation function
in Eq. (\ref{correl.func.101}) is calculated in two different ways.
The physical  or phenomenological part,  which is obtained in terms of the
hadronic parameters such as mass and decay constant inserting a complete set of
the states owing the same
quantum numbers as the interpolating current $j_{\mu\nu}$.
The theoretical or QCD  part,  which is calculated in terms of the QCD
parameters such as quark masses, quark condensates and quark-gluon coupling
constants etc. The correlation function in this part
is calculated in deep Euclidean region, $q^2\ll0$, via operator product
expansion (OPE).  The short distance effects are  calculated via the perturbation
theory, whereas the long distance contributions,
which are non-perturbative effects
are  parameterized in terms of  the quark-quark, gluon-gluon and quark-gluon
condensates.  The sum rules for the observables (masses and decay constants)
of the ground state $\chi_{Q2}(1P)$ meson are obtained
equating both representations of the correlation function,  isolating the ground
state and  applying Borel transformation to suppress the contribution of the higher
states and continuum through the dispersion relation.

To proceed first we calculate the phenomenological part. Inserting a complete set
of intermediate state,  $\chi_{Q2}(1P)$ to time ordering product in
Eq. (\ref{correl.func.101}), and performing integral over x we get:
\begin{eqnarray}\label{phen1}
\Pi _{\mu\nu,\alpha\beta}=\frac{{\langle}0\mid j _{\mu\nu}(0) \mid
\chi_{Q2}\rangle \langle \chi_{Q2}\mid j_{\alpha\beta}(0)\mid
 0\rangle}{m_{\chi_{Q2}}^2-q^2}
&+& \cdots,
\end{eqnarray}
 where $\cdots$ denotes the contribution of  the higher states and continuum.
From the above relation, it is clear that we need to know the matrix element,
$\langle 0 \mid j_{\mu\nu}(0)\mid \chi_{Q2}\rangle$, which can be
parameterized  in terms of the  decay constant, $f_{\chi_{Q2}}$:
\begin{eqnarray}\label{lep}
\langle 0 \mid j_{\mu\nu}(0)\mid \chi_{Q2}\rangle=f_{\chi_{Q2}}
m_{\chi_{Q2}}^3\varepsilon_{\mu\nu},
\end{eqnarray}
where $\varepsilon_{\mu\nu}$ is the polarization tensor of  $\chi_{Q2}$ meson.
Using Eq. (\ref{lep}) in Eq. (\ref{phen1}), we obtain the following final
representation of the correlation function
in phenomenological side:
\begin{eqnarray}\label{phen2}
\Pi _{\mu\nu,\alpha\beta}=\frac{f^2_{\chi_{Q2}}m_{\chi_{Q2}}^6}{m_{\chi_{Q2}}^2-q^2}
\left\{\frac{1}{2}(g_{\mu\alpha}g_{\nu\beta}+g_{\mu\beta}g_{\nu\alpha})\right\}+
\mbox{other structures}+...,
\end{eqnarray}
where, the only structure which contains a contribution of the tensor meson has
been kept. In calculations,
we have performed summation over the  polarization tensor  using
\begin{eqnarray}\label{polarizationt1}
\varepsilon_{\mu\nu}\varepsilon_{\alpha\beta}^*=\frac{1}{2}T_{\mu\alpha}T_{\nu\beta}+
\frac{1}{2}T_{\mu\beta}T_{\nu\alpha}
-\frac{1}{3}T_{\mu\nu}T_{\alpha\beta},
\end{eqnarray}
where,
\begin{eqnarray}\label{polarizationt2}
T_{\mu\nu}=-g_{\mu\nu}+\frac{q_\mu q_\nu}{m_{\chi_{Q2}}^2}.
\end{eqnarray}

The next step is to calculate the theoretical or  QCD side of the correlation
function in deep Euclidean region, $q^2\ll0$ . Using the explicit expression for
the tensor current presented in
 Eq. (\ref{tensorcurrent}) inside the correlation function shown in
Eq. (\ref{correl.func.101})  and contracting out all quark pairs using the
Wick's theorem, we obtain the following expression for the QCD side:
\begin{eqnarray}\label{correl.func.2}
\Pi _{\mu\nu,\alpha\beta}&=&\frac{i}{4}\int
d^{4}xe^{iq(x-y)} \Bigg\{Tr\left[S_Q(y-x)\gamma_\mu\olra{\cal D}_{\nu}(x)
\olra{\cal D}_{\beta}(y)S_Q(x-y)\gamma_\alpha\right]\nonumber \\
&&+\left[\beta\leftrightarrow\alpha\right]
+\left[\nu\leftrightarrow\mu\right]+\left[\beta\leftrightarrow\alpha,
\nu\leftrightarrow\mu\right]\Bigg\}.
\end{eqnarray}
To proceed we need to know  the heavy  quark propagator, $S_Q(x-y)$.
This propagator has been  calculated in \cite{22Balitsky}:

\bea\label{heavylightguy}
 S_Q (x-y)& =&  S_b^{free} (x-y) - i g_s \int \frac{d^4 k}{(2\pi)^4}
e^{-ik(x-y)} \int_0^1 dv \Bigg[\frac{\not\!k + m_Q}{( m_Q^2-k^2)^2}
G^{\mu\nu}[v(x-y)] \sigma_{\mu\nu} \nnb \\
\ar \frac{1}{m_Q^2-k^2} v (x_\mu-y_\mu)
G^{\mu\nu} \gamma_\nu \Bigg]~,
 \eea
where,
\bea\label{freeprop}
S^{free}_{Q}(x-y)
\es\frac{m_{Q}^{2}}{4\pi^{2}}\frac{K_{1}(m_{Q}\sqrt{-(x-y)^2})}{\sqrt{-(x-y)^2}}-i
\frac{m_{Q}^{2}(\not\!x-\not\!y)}{4\pi^{2}(x-y)^2}K_{2}(m_{Q}\sqrt{-(x-y)^2})~,
\eea
and $K_n(z)$ being the modified Bessel function of the second kind.
The next step is to use the expression of the heavy  propagators and perform the
derivatives with respect to x and y in  Eq. (\ref{correl.func.2}). After
setting  $y=0$,
the following final expression for the QCD side of the correlation function
in coordinate space is obtained:
\begin{eqnarray}\label{correl.func.3}
\Pi _{\mu\nu,\alpha\beta}=\frac{i}{64} \left(\frac{m_Q}{\pi}\right)^4
\int d^{4}xe^{iqx}\left\{\left[\Gamma_{\mu\nu,\alpha\beta}\right]+
\left[\beta\leftrightarrow\alpha\right]+\left[\nu\leftrightarrow\mu\right]+
\left[\beta\leftrightarrow\alpha,\nu\leftrightarrow\mu\right]\right\},
\end{eqnarray}
where,
\begin{eqnarray}\label{fonk}
\Gamma_{\mu\nu,\alpha\beta}&=&\Big\{-2 m_Q g_{\alpha\nu} g_{\beta\mu} {\cal
K}_1 {\cal K}_2 +
2 \Big( m_Q^2 x_\alpha x_\nu g_{\beta\mu} - g_{\alpha\nu} g_{\beta\mu} +
 g_{\alpha\mu} g_{\beta\nu} + g_{\alpha\beta} g_{\mu\nu}\Big) {\cal K}_2^2
\nonumber \\
&-& 2 m_Q^2  x_\alpha x_\nu g_{\beta\mu} {\cal K}_1 {\cal K}_3 -
2 m_Q g_{\alpha\nu} \Big(2 x_\beta x_\mu - x^2 g_{\beta\mu}\Big) {\cal K}_2
{\cal K}_3 \nonumber \\
&+& 2 m_Q^2  x_\alpha x_\nu \Big(2 x_\beta x_\mu - x^2 g_{\beta\mu}\Big)
{\cal K}_3^2  -
2 m_Q^2  x_\alpha x_\nu \Big(2 x_\beta x_\mu - x^2 g_{\beta\mu}\Big)
{\cal K}_2 {\cal K}_4 \Big\}\nonumber \\
&+&\mbox{nonperturbative contributions} ~,
\end{eqnarray}
and
\begin{eqnarray}
{\cal K}_n = {K_n(m_Q\sqrt{-x^2}) \over (\sqrt{-x^2})^n}~.
\end{eqnarray}

In the present work, we calculate the contributions of the heavy quark and gluon
condensates in nonperturbative part of the correlation function in QCD side. After a simple calculation we obtain for the heavy  quark condensate
(for the coefficient of the aforementioned structure)
\bea
\label{nolabel}
-{m_Q^3 \over 2(q^2-m_Q^2)} \lla \bar{Q}Q \rra~. \nnb
\eea

Using the well--known relation between the heavy quark and the gluon
condensates
\bea\label{nolabel}
m_Q \lla \bar{Q}Q \rra = - {1\over 12 \pi}
\lla {\alpha_s \over \pi} G^2 \rra~,\nnb
\eea
these two nonperturbative contributions can be written in terms of gluon
condensate contribution. Numerical analysis shows that, taking into account
quark condensates decreases gluon condensate contribution about 15\%.

Few words about the neglected dimension two operator in the operator
product expansion are in order. The term proportional to $1/q^2$
introduced in \cite{Chetyrkin0} is a phenomenological
parametrization of the higher order contributions to the perturbative
series. In other words, this term  can
appear when considering any types of  correlation functions where the
perturbative series are not zero. Obviously, this term  vanishes when
considering the difference of the correlators induced by vector and axial
vector currents, VV-AA in the chiral limit, $m_q=0$ (for more details see \cite{Narison}). In the 
present work, we neglect this term because we  work to  leading
order in $\alpha_s$.

 Now, we apply the Fourier transformation to the QCD side of the correlation
function to get its expression in momentum space. The next step is to select
the structure which gives contribution to the tensor state
from both sides of the correlation function,  equate the coefficient of the
selected structure from both sides and apply the Borel transformation to suppress
the contribution of the higher states and continuum. After lengthy calculations,
finally we obtain the following sum rules for the decay constant of the heavy
tensor quarkonia:
\begin{eqnarray}\label{sumrul}
f_{\chi_Q}^2e^{-m_{\chi_Q}^2/M^2} &=&\frac{N_c}{m_{\chi_Q}^6}\int_{4
m_Q^2}^{s_0} ds \int_1^\infty du \,
{e^{-s/M^2} [s-s(u)] \over 16 \pi^2 u^6}
\Big\{ - 2 m_Q^2 u^3 + [4 M^2 - s - s(u)] \nonumber \\
&-& 2 [4 M^2 - s - s(u)] u + [2 m_Q^2 + 4 M^2 - s - s(u)] u^2 \Big\} +
I(M^2) \lla {\alpha_s\over \pi} G^2 \rra,\nonumber \\
\end{eqnarray}
where,
\begin{eqnarray}\label{su}
s(u)=m_Q^2\left[u+\frac{1}{1-\frac{1}{u}}\right],
\end{eqnarray}
and the explicit expression of the function $I(M^2)$ is quite lengthy and
therefore we do not present it.

In the above sum rules, $M^2$ is the Borel mass parameter, $s_0$ is the
continuum threshold and $N_c=3$ is the   color factor. The mass of the heavy
tensor meson is also  obtained applying derivative with respect to
$-\frac{1}{M^2}$ to the both sides of the  sum rules for the decay constant
 and dividing by itself, i.e.,
\begin{eqnarray}\label{sumrul2}
m_{\chi_Q}^2&=&\Bigg(\int_{4
m_Q^2}^{s_0} ds \int_1^\infty du \,
{e^{-s/M^2} [s^2-s~s(u)] \over 16 \pi^2 u^6}
\Big\{ - 2 m_Q^2 u^3 + [4 M^2 - s - s(u)] \nonumber \\
&-& 2 [4 M^2 - s - s(u)] u + [2 m_Q^2 + 4 M^2 - s - s(u)] u^2 \Big\}\nonumber\\
&+&\int_{4 m_Q^2}^{s_0} ds \int_1^\infty du \,
{e^{-s/M^2} [s-s(u)] \over 16 \pi^2 u^6}
\Big\{ 4M^4(1-u)^2\Big\} - {d\over d(1/M^2)} I(M^2) \lla {\alpha_s \over
\pi} G^2 \rra
\Bigg)\nonumber\\&\times&\Bigg(\int_{4
m_Q^2}^{s_0} ds \int_1^\infty du \,
{e^{-s/M^2} [s-s(u)] \over 16 \pi^2 u^6}
\Big\{ - 2 m_Q^2 u^3 + [4 M^2 - s - s(u)] \nonumber \\
&-& 2 [4 M^2 - s - s(u)] u + [2 m_Q^2 + 4 M^2 - s - s(u)] u^2 \Big\}
+  I(M^2) \lla {\alpha_s \over\pi} G^2 \rra \Bigg)^{-1}.\nonumber \\
\end{eqnarray}

\section{Numerical analysis}

In this section, we numerically analyze  the sum rules for the mass and decay
constant of the ground state tensor quarkonia.  Some input  parameters entering
the sum rules are quark masses, $m_b=(4.8\pm 0.1)~GeV$, $m_c=(1.46 \pm 0.05)~GeV$
\cite{colangelo} and gluon condensate, $\langle 0|\frac{1}{\pi}\alpha_{s}G^{2}|0\rangle=(0.012
\pm 0.004)~ GeV^{4}$. It should be noted that recent analysis of experimental data leads to the
$\langle 0|\frac{1}{\pi}\alpha_{s}G^{2}|0\rangle=(0.005
\pm 0.004)~ GeV^{4}$ for the gluon condensate  \cite{Ioffeb}. For conservative estimation in numerical analysis, we also take into account the value
of gluon condensate $\langle 0|\frac{1}{\pi}\alpha_{s}G^{2}|0\rangle=(2.16\pm0.38)\times 10^{-2}~GeV^4$ which follows
from sum rules of $e^{+}e^{-} \rightarrow I=1~ hadrons$ \cite{12} and heavy quarkonia \cite{13,14,15}. Few words about quark mass are in order. The aforementioned masses are the pole masses for the quarks. Using the four loop results for
the vacuum polarization operator in \cite{Chetyrkin},  the running masses of the charm and
beauty quarks, $m_c(3GeV)=(0.986\pm0.013)~GeV$ and $m_b(m_b)=(4.163\pm0.016)~GeV$ are obtained. These   improved values of the running masses of charm and
beauty quarks as well as  wide range of gluon
condensate are  used in numerical calculations.  To obtain more reliable results for the mass and decay constant of the heavy tensor meson, we will also  take into account
 a more realistic error coming from
 the range spanned by the pole and
running quark masses as well as the range for the value of the gluon condensate.

 From the sum
rules for the decay constant and mass it is clear that they contain also two
auxiliary parameters, continuum threshold $s_0$
and Borel mass parameter $M^2$. The standard criteria in QCD sum rules is that
the physical quantities should be independent of these mathematical objects,
so we should look for working regions for these parameters
at which the masses and decay constants  practically remain unchanged. To determine the working
region for the Borel mass parameter
the  procedure is as follows: the lower limit of $M^2$ is obtained requiring
that the higher states and continuum contributions constitute, say,
30\% of the total dispersion integral. The upper limit of $M^2$ is chosen demanding
that the sum rules for the decay constants and masses should be convergent, i.e.,
contribution of the operators with higher dimensions is small. As a result, we
choose the  regions: $ 8~ GeV^2 \leq
M^2_ {\chi_{b2}}\leq 20~ GeV^2 $ and $ 4~ GeV^2 \leq
M^2_ {\chi_{c2}}\leq 7~ GeV^2 $ for the Borel mass parameter. The continuum threshold $s_0$ is not completely arbitrary but it is  correlated to the energy of the first exited
state with quantum numbers of the interpolating current. Our numerical results are in consistency with this point and   show that  in the interval $(m_{\chi_{Q2}}+0.4)^2\leq
s_0^{\chi_{Q2}}\leq (m_{\chi_{Q2}}+0.7)^2$, the results are practically insensitive to the variation of this parameter.
Here we would like to make the following
remark. It is shown in \cite{Lucha2} that the continuum
threshold $s_{0}$  can depend
on the Borel mass parameter. Therefore, the standard criteria,
namely, weak dependence of the results on variation of the auxiliary parameters
does not provide us realistic errors, and in fact the actual
error should be large. Following \cite{Lucha2}, in the present work we will
add also  the systematic errors to the numerical values.

Our numerical analysis on the masses and decay constants leads to the results presented in Table 1.
\begin{table}[h] \centering
 \begin{tabular}{|c||c|c|c|c|} \hline &
 Present Work&Experiment \cite{Amsler}
\\\cline{1-3}\hline\hline
$m_{\chi_{b2}}$& $(9.90\pm2.48)~GeV $ &$(9.91221\pm0.00057)~GeV$\\
\cline{1-3} $m_{\chi_{c2}}$& $(3.47\pm0.95)~GeV $ &$(3.55620\pm0.00009)~GeV$\\
\cline{1-3} $f_{\chi_{b2}}$& $0.0122\pm0.0072 $ &-\\
\cline{1-3}$f_{\chi_{c2}}$& $0.0111\pm0.0062 $ &-\\
\cline{1-3}\hline\hline
\end{tabular}
\vspace{0.8cm} \caption{Values for the masses and decay constants of the tensor
mesons $\chi_{Q_2}$.} \label{tab:27}
\end{table}
The quoted errors in our predictions are due to the variations in the continuum
threshold and Borel parameter, uncertainties in quark masses and wide range of the gluon condensates presented at the beginning of this section  as well as the systematic errors. The results presented in Table 1 show a good consistency between our predictions
and  the experimental values  \cite{Amsler} on the masses of the ground state
heavy, $\chi_{b2}$ and, $\chi_{c2}$ tensor mesons.
 Our predictions on the
decay constants can be verified in the future experiments.

\section{Acknowledgment}
We thank  A. Ozpineci for his useful discussions.


\begin{thebibliography}{99}

\bibitem{Amsler} C. Amsler et al., (Particle Data Group),
Phys. Lett. B 667 1 (2008).

\bibitem{Ecklund} K. M. Ecklund et al., [CLEO Collaboration],
Phys. Rev. D 78 :091501 (2008).

\bibitem{svz} M. A. Shifman, A. I. Vainshtein, V. I. Zakharov,
Nucl. Phys. B 147 (1979) 385.

\bibitem{colangelo} P. Colangelo, A. Khodjamirian, in "At the
Frontier of Particle Physics/Handbook of QCD", edited by
M. Shifman (World Scientific, Singapore, 2001), Vol. 3, p. 1495.

\bibitem{kazem1} T. M. Aliev, K. Azizi, V. Bashiry,
J. Phys. G 37, 025001 (2010), arXiv:0909.2412 [hep-ph].

\bibitem{aliev} T. M. Aliev, M. A. Shifman,
Phys. Lett. B 112  , 401 (1982).

\bibitem{kazem2} T. M. Aliev, K. Azizi, M. Savci,
arXiv:0909.2413 [hep-ph].

\bibitem{22Balitsky} I. I. Balitsky,  V. M. Braun,
Nucl. Phys. B 311 (1989) 541.

\bibitem{Chetyrkin0}  K. G. Chetyrkin, S. Nasrison, V. I. Zakharov, Nucl. Phys. B 550 (1999) 353.
\bibitem{Narison}  S. Narison,   V. I. Zakharov, Phys. Lett. B 522 (2001) 266.
\bibitem{Ioffeb} B. L. Ioffe,  Prog. Part. Nucl. Phys. 56 (2006) 232.
\bibitem{12}  S. Narison,  Phys. Lett. B 387 (1996) 162.
\bibitem{13} J. S. Bell, R. Bertlmann, Nucl. Phys. B 177 (1981) 218
\bibitem{14} J. S. Bell, R. Bertlmann, Nucl. Phys.  B 187 (1981) 285.
\bibitem{15} F. J. Yndurain, arXiv: hep-ph/9903457 (1999).
\bibitem{Chetyrkin} K. G. Chetyrkin et. al, Phys. Rev. D 80, 074010 (2009).

\bibitem{Lucha2} W. Lucha, D. Melikhov,  S. Simula,
Phys. Rev. D 79, 0960011 (2009).

\end{thebibliography}
\end{document}